\documentstyle[aps,epsf]{revtex}    

\begin{document}

\baselineskip=25pt

\language=1
\begin{center}{

\Large
Collective effects in the 2D lattices of the magnetic nanoparticles.

\vskip 24pt
{\large A.A.Fraerman, S.A.Gusev, L.A.Mazo, I.M.Nefedov, Ju.N.Nozdrin,
M.V.Sapozhnikov, L.V.Suhodoev\\

\vskip 12pt
Russian Academy of Science\\
Institute for Physics of Microstructures\\
\vskip -3pt GSP-105, Nizhny Novgorod, 603600, Russia\\
E-mail: msap@ipm.sci-nnov.ru}}
\end{center}
\vskip 2mm


\vskip 20mm

\begin{abstract}
The work is devoted to investigation of the collective behavior of the
regular rectangular lattices of the $Ni_3 Fe$ nanoparticles
caused by the dipole-dipole interaction between them. The samples was
prepared by the method of the electron lithography and consist of about
$10^5$ particles of $\sim$50 nm size. The magnetization curves were
investigated by the Hall magnetometry for the different
orientation of the external magnetic field at $4.2^\circ$K and
$77^\circ$K. The results points on the collective behavior of the system.
The observed peculiarities we connect with the quasi-one-dimensional
behavior of the system and the formation of the solitons in the system.
\end{abstract}

The physical properties of arrays of small ferromagnetic particles
continue to be an active area of fundamental and theoretical research.
The intensive study of those systems was stimulated, in part, by various
applications \cite{abstr_MMM96}. One of the most interesting effect is
the supermagnetic ordering of the system of the interacting ultrafine
magnetic particles \cite{morup_83}. In the case of the dipole-dipole
interaction, the critical temperature $T_c$ of the transition in the
superparamagnetic state is

$$ T_c\sim M^2 / R^3, $$

where $M$ is the magnetization of the particles, $R$ is the
interparticle distance. For example, if the radius of the
transition-metal particle is 10nm, $R\sim 100$nm, we have for $T_c\sim
100^\circ$K. Let`s mark also, that the type of the long range order in
the 2D lattice of the dipole-dipole interacted particles strongly
dependences on the symmetry of the lattice due to the long range and
anisotropic character of this interaction. It is known
\cite{rozenbaum_91r,rozenbaum_91r_1}, that the ground state of the
classical 3D dipoles on the rectangular lattice is of the
antiferromagnetic type, on the square one is microvortixes.  The phase of
the dipole glass can be the ground state for the random array

For the experimental observation of the  supermagnetic transition we must
have a regular lattice of the single domain ferromagnetic particles.
Moreover the switching time \cite{wernsdofer_96} for this particles at
$T\sim T_c$ are found to be less than the time it takes for the
measurements. As it is known, the switching time exponentially decreases
as the particle volume and coercetive field decreases.

Advances in the nanotechnology make possible the experimental
observation of the collective behavior of the small ultrafine
ferromagnetic particles. In \cite{hauschild_98} the experimental observation of
the supermagnetic transition in the linear self-assembling mesoscopic
$Fe$ particle arrays is presented. In \cite{saguwara_97} the dipolar
supermagnetism in the monolayer nanostripes of $Fe$ on the vicinal
$W$(110) surfaces is reported. By contrast with
\cite{hauschild_98,saguwara_97} in this letter we carried out the regular
2D rectangular lattices of the nanometer-scale magnetic particles created
by the electron lithography method. We observed the unusual magnetic
hysteresis for magnetization in the perpendicular direction and
investigated main properties of this collective phenomena.  It is
possible that this effect is determined by the temperature induced
nonuniform states which are typical for quasi-one-dimensional systems.

The rectangular lattices of the magnetic nanoparticles are prepared by
the electron lithography method from the $Ni_3 Fe$ films. The double-layer
mask containing the $C_{60}$ film as a negative electron resist and the
$Ti$ film as a transmitting layer was used. $Ni_3 Fe$ and $Ti$ layers are
prepared by the laser deposition method at the room temperature. The
$C_{60}$ films were prepared by the sublimation  of a the $C_{60}$ -
powder at the temperature $350^\circ C$ in the vertical reactor with hot
walls. The thickness of the masking layers was 20 and 30 nm, accordingly.
The thickness of the magnetic layers is varied from 25 nm (samples 2,3)
to 45 nm (sample 1). The $C_{60}$ films is exposed in the JEM - 2000 EXII
electron microscope. The parameters of the exposition were the following:
the accelerating voltage 200 kV, the exposure dose 0.05 $\div$ $0.1
C/cm^2$.  The diameter of the electron beam, according to our
estimations, was 300-700 $\AA$. The electron beam irradiation of the
$C_{60}$ films reduces the solubility of the fullerenes in the organic
solvents.  The most likely reasons of the changes of the solubility are
the electron induced polymerization of the $C_{60}$ molecules accompanied
partially graphitization ones \cite{gusev_97}. The exposed films were
developed in the toluene during 1 min, the $Ti$ etching was carried out
by a plasma ething method in the $CF_2 Cl_2$ atmosphere, the $Ni$ etching
by the ion milling at the $Ar$ atmosphere. The SEM-images of the one of
the samples are shown on the Fig. 1.

The height of the particles is determined by the thickness of the $Ni_3
Fe$ layer. Their diameter dependences, in the main, on the exposed time.
The parameters of the investigated samples are summarized in the Table
below. There a and b are the lattice parameters, h is the high of the
particles and d is their diameter. The total number of the particles is
equal to $10^5$.  \vskip 10mm

\begin{figure}[th]
\centerline{
\epsfxsize=7cm
\epsffile{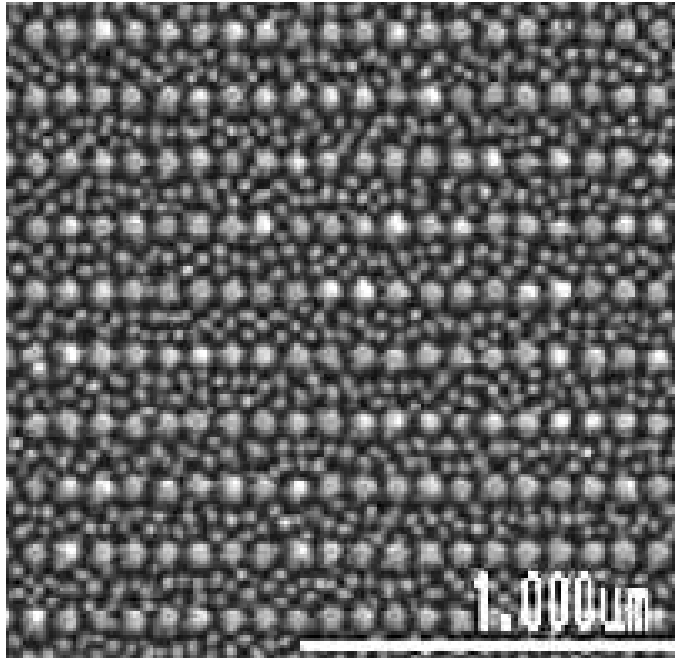}}
\caption[b]{

The SEM-image of the sample 1 (See the table 1). The lattice of the 40-50
nm particles is visible with the background of the 10 nm roughnesses of
the sublayer.} \end{figure}

\begin{center}
\begin{tabular}{c|c|c|c|c}
\hline
&&&&\\
N&a(nm)&b(nm)&h(nm)&d(nm)\\
&&&&\\
\hline
&&&&\\
1&90&180&45&50\\
&&&&\\
2&120&240&25&80\\
&&&&\\
3&120&120&25&70\\
&&&&\\
\hline
\end{tabular}
\end{center}

\vskip 10mm

The measurements of the magnetic properties were provide using the
commercial magnetometer based on the Hall response in a semiconductor
(InSb). The widths of the current and the voltage probes are 0.2mm
and 0.06mm and the thickness of the semiconductor layer is 10$\mu$m. The
lattice of the particles is located in the active area of one of the Hall
crosses. The difference in the Hall voltage between this
sample cross and the closely spaced reference is measured using the
bridge circuit \cite{kent_94}.
 Withe the bridge properly balanced, the
resulting output voltage is proportional to the sample contribution to
the magnetic induction. This contribution can then be calculated so that
$\Delta B=V/RI$ where R is the Hall coefficient and I is the measured
current. Typically we use the dc current of 50 mA. The large Hall
response in the combination with the good coupling of the small samples
to the device results in the excellent spin sensitivity (ratio
signal/noise is approximately 100). The sensor works over a large range
of the magnetic field and temperature.  We investigate the magnetic
properties of the samples by measuring the perpendicular magnetization as
a function of the direction and magnitude of the applied field. The
particles have the polycrystal structure (this was determined by the
X-ray and Selected area diffraction) and do not have the anisotropy of
the form in the plane of the system. In this case the difference of the
magnetization curves for the different orientation of the external
magnetic field
to the structure lattice is a positive attribute of the
collective behavior of the system.  As the used method allows to measure
only the z-component of the magnetization, we provide our investigation
with the thee orientation of the external magnetic field:  1) the field
is perpendicular to the sample plane ($\theta=0^{\circ}$); 2)
the field is directed at 45 deg.  to the sample plane along the short
side of the rectangle cell ($\theta=45^{\circ}$, $\phi=0^{\circ}$); 3)
the field is directed at 45 deg. to the sample plane along the long side
of the rectangle cell ($\theta=45^{\circ}$, $\phi=90^{\circ}$). The
results of this measurements for $T=4.2 K$ are represented on the Figs.
2, 3, 4  accordingly. The difference in the magnetization curves
indicates the collective behavior of the system, which is the result of
the dipole-dipole interaction between particles.
The hysteresis if the field directed at $\theta=45^{\circ}$,
$\phi=0^{\circ}$ (Fig. 3) is the attribute of the easy axis of the
magnetization which is directed along the short side of the rectangle
cell.  The remanent magnetization is absent in this case. The existence
of such anisotropy in the dipole system was theoretically predicted
\cite{rozenbaum_91r_1}.   The magnetization curves if the field directed at
$\theta=0^{\circ}$  or $\theta=45^{\circ}$, $\phi=90^{\circ}$ (Figs. 2,
4) are qualitatively similar. They have hysteresis in the weak magnetic
field with the remanent magnetization which is approximately 5 \% of the
saturation magnetization. The saturation magnetization of the first
sample is 35G.

\begin{figure}[th]
\centerline{
\epsfxsize=7cm
\epsffile{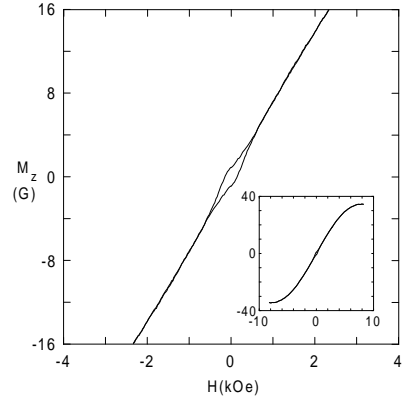}}
\caption[b]{
The dependence of $M_z$ on the magnetic field with $\theta=0^{\circ}$.
The whole magnetization curve is shown on the casing-in.} \end{figure}

\begin{figure}[th]
\centerline{
\epsfxsize=7cm
\epsffile{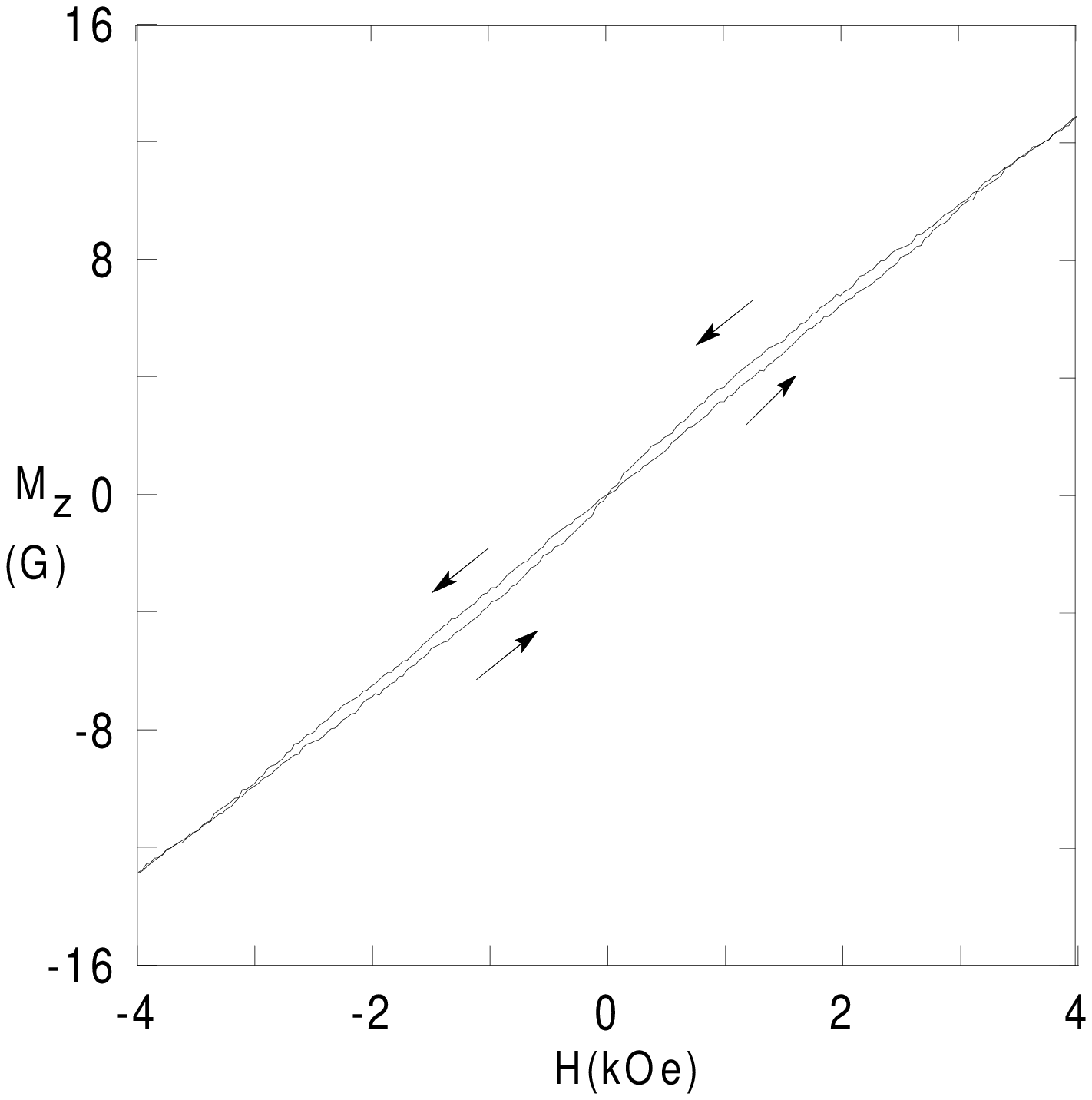}}
\caption[b]{
The dependence of $M_z$ on the magnetic field with $\theta=45^{\circ}$,
$\phi=0^{\circ}$.} \end{figure}

\begin{figure}[th]
\centerline{
\epsfxsize=7cm
\epsffile{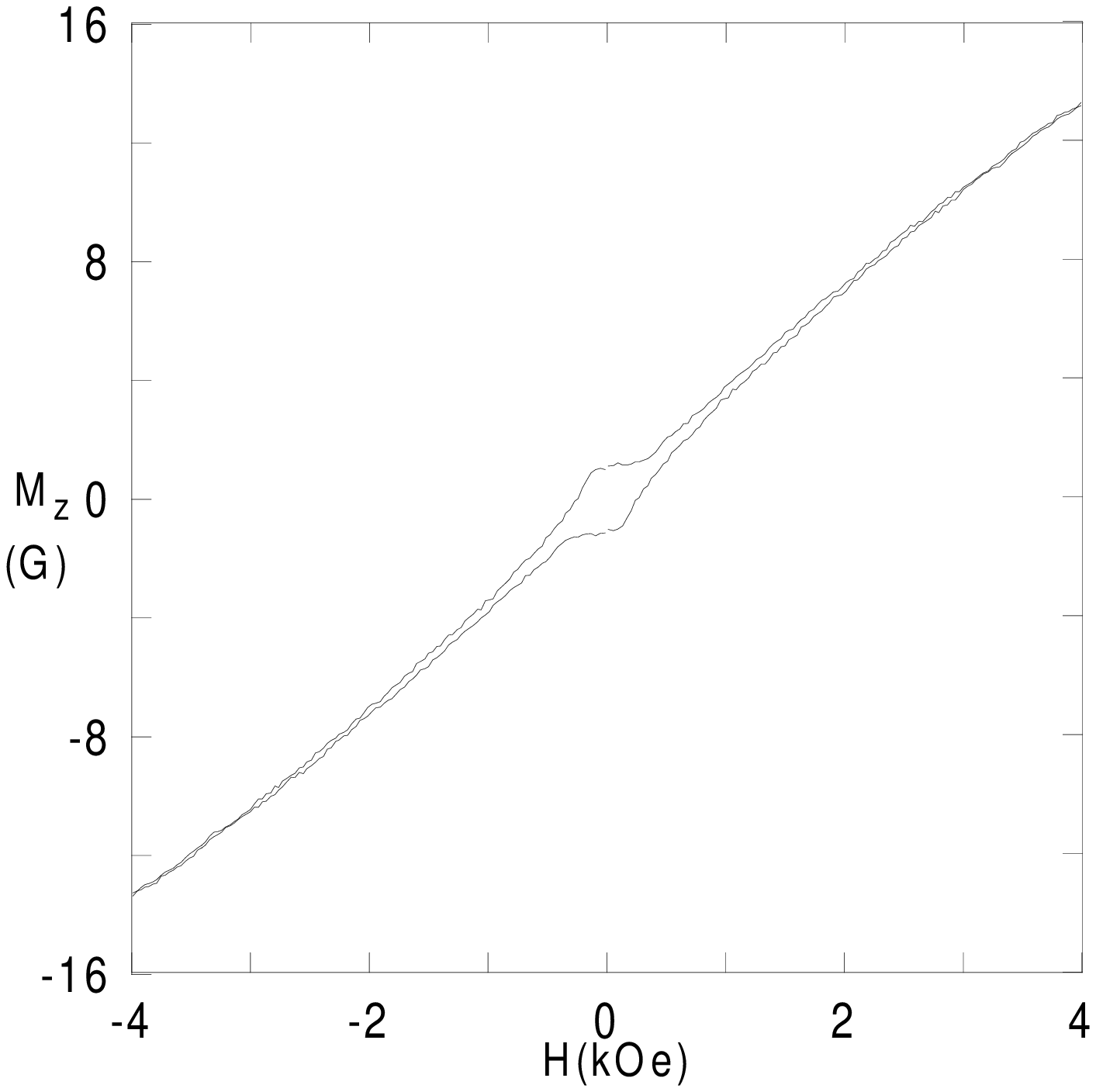}}
\caption[b]{
The dependence of $M_z$ on the magnetic field with $\theta=45^{\circ}$,
$\phi=90^{\circ}$.} \end{figure}
The magnetization curves for the second sample is qualitatively similar
the those of the first sample, although the samples themselves has the
difference shape of the particles. The saturation magnetization of the
second one is 18G.

The measurements of the magnetization curves for the system with the
square lattice (sample 3), shows the absence of the anisotropy while
magnetizing by field directed at $\theta=45^{\circ}$, $\phi=0^{\circ}$ or
at $\theta=45^{\circ}$, $\phi=90^{\circ}$. This fact approves the
isotropy of the single particle in the plane of the system. The
hysteresis is absent for any direction of the external field, the
remanent magnetization is equal to 0 also. The saturation magnetization
of the third sample is 20G.

All results: the existence of the anisotropy axis in the plane of the
sample with the rectangular lattice and its absence for the sample with
the square lattice were predicted earlier \cite{rozenbaum_91r} and were
principally expected. As for the hysteresis of the magnetization curves
and the remanent magnetization for the sample with the rectangular
lattice if the external magnetic field direction is $\theta=0^{\circ}$ or
$\theta=45^{\circ}$, $\phi=90^{\circ}$, their existence were unexpected.
The effect can not be explained by the single particle properties. In
this case the remanent magnetization must exist for the every direction
of the magnetic field. Besides it must be observed for the sample with
the square lattice in this case. Evidently, the existence of the
remanent magnetization is the exhibition of the collective properties of
the dipole system with the rectangular lattice.

The first sample was investigated at T=77K also. The hysteresis with the
field directed at $\theta=45^{\circ}$, $\phi=0^{\circ}$ did not be
observed. The hysteresis with the field directed at $\theta=0^{\circ}$ or
$\theta=45^{\circ}$, $\phi=90^{\circ}$  was qualitatively changed (Fig.
5). This fact directly indicates the dependence of the collective
properties on the temperature.

\begin{figure}[th]
\centerline{
\epsfxsize=7cm
\epsffile{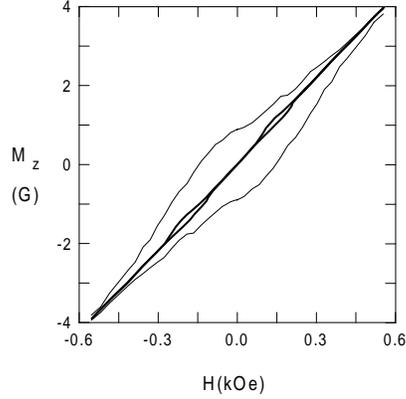}}
\caption[b]{
The changing of the magnetization curve hysteresis with the temperature:
the thin line for $4.2^{\circ}K$,  the thick one for $77^{\circ}K$. H is
directed at $\theta=0^{\circ}$} \end{figure}

Our samples can be considered as a lattices of the interacting
single-domain ferromagnetic particles. Model of the 3D classical dipoles
is a good approximation for those systems. Energy of the system has the
form

$$
E=\frac{1}{2}\sum_{{\bf r} \neq {\bf r`} }D_{ik}({\bf r}-{\bf r`})
M_i({\bf r})M_k({\bf r`})+\frac{K}{2}\sum_{{\bf r}}M_z^2({\bf r})-
H_i\sum_{{\bf r}}M_i({\bf r}),
$$
$$
D_{ik}=\frac{\delta_{ik}}{r^3}-\frac{3r_i r_k}{r^5}.
$$

The numerical simulation were provide for such rectangular (1:2)  and
square lattices of the dipoles with sizes 20$\times$50 and
30$\times$30 particles. The system of relaxation Landau-Lifshitz
equations were solved with the effective field on the particle equal to
the sum of the dipole field of the other particles. The equations were
solved with the different values of the single particle anisotropy
(the anisotropy axis is perpendicular to the lattice, as our particles
are isotropic in the plane of the system). The relaxation scheme was as
follows: at first the randomly directed dipoles relaxed in the
external magnetic field which was approximately 1/3 of the saturation
field.  After that the field was switched off and the following
relaxation without external field was performed. The directions of the
initial field to the system lattices were similar to those in
the experiment.  The following results were obtained.

In the case of the square lattice the system relaxes to the microvortex
state \cite{rozenbaum_91r} with $M_z=0$ independently on the initial
external field orientation.

As for the rectangular lattices, if the initial field was oriented along
$\theta=45^{\circ}$, $\phi=0^{\circ}$ direction, the final state had the
uniform magnetization along the short side of the rectangular. If the
initial field has $\theta=0^{\circ}$  or $\theta=45^{\circ}$,
$\phi=90^{\circ}$ orientation, the system relaxes to the state with the
solitons in the chains of the dipoles. They have the antiferromagnetic
core withe the dipjles oriented perpendicular to chains and
ferromagnetic tails in which dipoles oriented alog the chains in the
opposite directions.  The antiferromagnetic core is narrow (it takes
only 3-7 cells in the zero magnetic field), but it can have the magnetic
moment which value and direction depends on the value of the single
particle anisotropy.
If the
single particle anisotropy is absent, the soliton lies in the plane of
the lattice and does not have magnetic moment. With the increase of the
single particle anisotropy the soliton obtains the magnetic moment
directed along z-axis. For example, the z-component of the soliton
magnetic moment is approximately 0.8 of the single particle magnetic
moment, if K=3.

If the anisotropy value is greater than a certain critical one, the system
relaxes in the state uniformly magnetized along z-direction
independently on the lattice symmetry and initial direction of the
external field.

It is necessary the length of the dipoles chains to be at list of
approximately 30 dipoles for the stability of the soliton. If the size of
the system is less, the soliton, even appearing in the initial field, at
zero field ran out from the chain. Evidently, in this case the
force attracting the soliton to the boundary is greater than the pinning
force on the discrete lattice. The magnetization reverse of the soliton
takes place in the external field mach less than the saturation field.

Let us notice especially, that in the case then magnetic moment of the
soliton is directed along z-axis, e.i. single particle anisotropy is
z-"easy axis", the whole system has the anisotropy of the "easy axis"
lying in the plane of the system along the chains of the dipoles (along
x-axis), and in the main state dipoles in the chains directed along the
chain, but the chains themselves antiferromagnetically ordered.

Using the results of the numerical simulation we suggested the following
mechanism of the appearing of the remanent magnetization of our samples
with the rectangular lattice. Due to quasi 1D character of the system
in the fields near to the saturation the soliton energy is small and
their thermoinduced formation is possible. With the decrease of the
external field the their energy  increases, the width narrows. It leads
to their pinning on the discrete lattice. So the remanent magnetization
is caused by the existence of the "frozen" solitons, which have formed in
the large field. The necessary condition of the z-component of the
remanent magnetization of the soliton is the single particle anisotropy.
As the particles of the second sample have the disk form, there must be
another mechanism of the appearance of the "easy axis" single particle
anisotropy (surface, magnetostriction, etc.). The additional experiments
are to be carried out to check its existence.

\section*{Acknowledgment.}
The authors acknowledge many helpful discussions with A.A.Andronov. The
work was supported by the Russian Foundation for Fundamental Research (N
59-02-05388 ).

\end{document}